\documentclass[11pt]{article}
\usepackage{graphicx}
\usepackage{cite,graphicx,subfigure,rotating,epsfig}

\newcommand{\BABARPubYear}    {05}

\newcommand{\BABARConfNumber} {002}
\newcommand{\SLACPubNumber} {11309}
\newcommand{\LANLNumber} {0000}

\input babarsym
\def\Bztokkkl	{\ensuremath{\Bz \to K^+K^-\KL}}
\def\Bztokkks	{\ensuremath{B^0 \to K^+K^-\KS}}
\def\Bztojpsikl     {\ensuremath{B^0 \to \jpsi\KL}}
\def\kkks	{\ensuremath{\Kp\Km\KS}}
\def\kkkl	{\ensuremath{\Kp\Km\KL}}
\def\phiks	{\ensuremath{B^0 \to \phi\KS}}
\def\phikl	{\ensuremath{B^0 \to \phi\KL}}

\def\Bflav {\ensuremath{B_{\mbox{flav}}}\xspace}
\def\Btag {\ensuremath{B_{\mbox{tag}}}\xspace}
\def\Bcp {\ensuremath{B_{CP}}\xspace}
\def\finalscb{\ensuremath{0.07\pm 0.28\, (\mbox{\small stat})^{+0.11}_{-0.12}
(\mbox{\small syst})} \xspace}
\def\finalccb{\ensuremath{0.54\pm 0.22\, (\mbox{\small stat}) ^{+0.08}_{-0.09}, (\mbox{\small syst})} \xspace}
\def\finalstwob  {\ensuremath{0.09 \pm 0.33\, (\mbox{\small stat})  
		 ^{+0.13}_{-0.14} (\mbox{\small syst}) \pm 0.10 (\mbox{\small syst \CP-cont})}}

\long\def\inst#1{\par\nobreak\kern 4pt\nobreak
    {\it #1}\par\vskip 10pt plus 3pt minus 3pt}

\begin{document}
{\pagestyle{empty}

\begin{flushright}
\babar-CONF-\BABARPubYear/\BABARConfNumber \\
SLAC-PUB-\SLACPubNumber \\
hep-ex/\LANLNumber \\
July 2005 \\
\end{flushright}

\par\vskip 5cm

\begin{center}
\Large \bf  Measurement of Time-dependent {\boldmath $\CP$}-Violating
Asymmetries in {\boldmath \Bztokkkl } Decays
\end{center}
\bigskip

\begin{center}
\large The \babar\ Collaboration\\
\mbox{ }\\
\today
\end{center}
\bigskip \bigskip

\large
We present preliminary measurements of the \CP\ asymmetry parameters and \CP\ content in
$\Bz \to \Kp\Km\KL$\ decays, with \phikl\ events excluded.
In a sample of 227~M \BB\ pairs collected by the \babar\ detector at the \pep2\
\BF\ at SLAC, we find the \CP\ parameters to be
        $$S      =       0.07 \pm 0.28 (\mbox{\small stat}) ^{+0.11}_{-0.12} (\mbox{\small syst}) $$	
        $$C      =       0.54 \pm 0.22 (\mbox{\small stat}) ^{+0.08}_{-0.09} (\mbox{\small syst})$$
where the first error is statistical and the second is systematic.
Estimating the fraction of \CP -odd final states from angular moments analysis in the \kkks\ \CP -conjugate 
final state, 
$
f_{odd} (\kkkl) ~=~ 0.92 \pm 0.07~(\mbox{stat}) \pm 0.06~(\mbox{syst})
$, 
we determine
$$
\stwob_{\mbox{eff}}~=~\finalstwob
$$
where the last error is due to uncertainty on the \CP\ content.

\vfill
\begin{center}
Contributed to the 
XXII$^{\rm st}$ International Symposium on Lepton and Photon Interactions at High~Energies, 6/30 --- 7/5/2005, Uppsala, Sweden
\end{center}

\vspace{1.5cm}
\begin{center}
{\em Stanford Linear Accelerator Center, Stanford University, 
Stanford, CA 94309} \\ \vspace{0.1cm}\hrule\vspace{0.1cm}
Work supported in part by Department of Energy contract DE-AC03-76SF00515.
\end{center}

\newpage
} 

\begin{center}
\small

The \babar\ Collaboration,
\bigskip

B.~Aubert,
R.~Barate,
D.~Boutigny,
F.~Couderc,
Y.~Karyotakis,
J.~P.~Lees,
V.~Poireau,
V.~Tisserand,
A.~Zghiche
\inst{Laboratoire de Physique des Particules, F-74941 Annecy-le-Vieux, France }
E.~Grauges
\inst{IFAE, Universitat Autonoma de Barcelona, E-08193 Bellaterra, Barcelona, Spain }
% ...

\end{center}\newpage

\section{INTRODUCTION}
\label{sec:Introduction}
In the Standard Model (SM), \CP violation arises from a single
complex phase in the Cabibbo--Kobayashi--Maskawa (CKM) quark-mixing
matrix~\cite{ckm}. 
Decays of \B\ mesons into charmless hadronic final states with three 
kaons are dominated by $b \to s\bar s s$ 
gluonic penguin amplitudes, with smaller contributions from electroweak
penguins, while other SM amplitudes are suppressed by CKM factors~\cite{one}.
The time-dependent \CP-asymmetry is obtained by measuring the proper time difference 
$\deltat = t_{CP} - t_{\mbox{tag}}$ between
a fully reconstructed neutral $B$ meson (\Bcp) in the final state \kkkl, and the partially reconstructed
recoil $B$ meson (\Btag). The decay products of \Btag\ provide evidence that it decayed either as \Bz\ 
or \Bzb\ (flavor tag).
The decay rate ${\mbox{f}}_+({\mbox{f}}_-)$ when the tagging meson is a $\Bz (\Bzb)$
is given by
\begin{eqnarray}
{\mbox{f}}_\pm(\, \deltat)& = &{\frac{{\mbox{e}}^{{- \left| \deltat
\right|}/\tau_{\Bz} }}{4\tau_{\Bz}}}  \, [
\ 1 \hbox to 0cm{}
\pm
S \sin{( \deltamd  \deltat )}
\mp
\,C  \cos{( \deltamd  \deltat) }   ],
\label{eq::timedist}
\end{eqnarray}
where $\tau_{\Bz}$ is the neutral $B$ meson mean lifetime and
\deltamd is the \Bz--\Bzb oscillation frequency.
The parameters $C$ and $S$ describe the magnitude of \CP violation in the decay and
in the interference between decay and mixing, respectively.
The time-dependent \CP -violating asymmetry is defined as
$A_{\CP} \equiv ({\mbox{f}}_+  -  {\mbox{f}}_- )/
({\mbox{f}}_+ + {\mbox{f}}_- )$.
In the SM, we expect $C=0$ because there is only one decay mechanism and direct \CP\ violation
requires amplitudes with different phases.
Neglecting CKM-suppressed contributions and assuming that 
\kkkl\ decay proceeds through an S-wave, leading to a \CP-odd final state,
the time-dependent \CP-violating parameter $S$ in this decay
and $\Bz\to \jpsi K^0$ are both equal to the same parameter
$\sin 2\beta$~\cite{grossman}, where the latter decay is dominated by
tree diagrams. Since many scenarios of physics beyond the SM introduce
additional diagrams with heavy particles in the penguin loops and
corresponding new phases, comparison of \CP-violating observables with SM
expectations is a sensitive probe for new physics~\cite{Ciuchini:2002uv}.
Measurements of \stwob in $B$ decays to charmonium such as $\Bz\to J/\psi \KS$
have been reported by the \babar~\cite{sin2bnewbabar} and Belle~\cite{sin2bnewbelle}
collaborations, and the world average for \stwob ($0.736\pm 0.049$~\cite{pdg}) is in
good agreement with SM expectations~\cite{fits}. A deviation from this value in the case
of loop-dominated channels might signal the presence of physics beyond the SM. 

Measurements of the \CP\ asymmetry in the decays \phiks\ and \phikl\ currently have large 
statistical uncertainties~\cite{Abe:2003yt,Aubert:2004ii}. 
More accurate \CP\ asymmetry measurements have been performed in the final state \kkks, 
(excluding \phiks)~\cite{kkksconf},
which has a branching fraction several times larger than the resonant modes,
The \CP content of the final state,
which is {\it a priori} unknown, is estimated using an angular-moment analysis.

In this document we report preliminary measurements of the time dependent \CP\ asymmetry 
in the \CP\ conjugate state \Bztokkkl.

\section{THE \babar\ DETECTOR AND DATASET}
\label{sec:babar}
This measurement is  based on a sample of approximately
227 million $B\bar{B}$ pairs collected at the $\Upsilon(4S)$ resonance
with the \babar\ detector~\cite{Aubert:2001tu} at the \pep2\ asymmetric-energy
\epem storage ring~\cite{pep} located at the Stanford Linear Accelerator Center.
The \babar\ detector is fully described elsewhere~\cite{Aubert:2001tu}.
The detector systems used in this analysis are a charged-particle
tracking system consisting of a five-layer silicon vertex tracker
(SVT) and a 40-layer drift chamber (DCH) surrounded by a 1.5-T solenoidal
magnet with an instrumented flux return (IFR), an electromagnetic calorimeter
(EMC) composed of 6580 CsI(Tl) crystals, and a detector of internally reflected
Cherenkov light (DIRC) providing excellent charged $K-\pi$ identification up to
a momentum of 4.5~\gevc, which is the relevant momentum range for this analysis.

\section{EVENT RECONSTRUCTION}
\label{sec:Reco}
The \Bztokkkl{} candidate ($B_{CP}$) is reconstructed by combining a pair of oppositely charged tracks 
extrapolated from a common vertex and a \KL\ candidate. 
For the charged tracks we require at least 12 measured drift-chamber 
coordinates and a minimum transverse momentum of
0.1~\gevc. The tracks must also originate within $\pm 10$~cm along the 
beam axis and 1.5~cm in the transverse plane, with respect to the nominal beam spot.
Charged kaons are distinguished from pion
and proton tracks via a requirement on a likelihood ratio
that combines \dedx information
from the SVT and the DCH
for tracks with momentum $p < 0.7 \gevc$.
For tracks with higher $p$, \dedx in the DCH and the
Cherenkov angle and the number of photons as measured by the
DIRC are used in the likelihood. 
These particle identification criteria limit the rate of pion
misidentification as a kaon to less than 2\%,
with an efficiency of 70\%.

We identify a \KL candidate as in the \babar\ analysis of the decay
$\Bz\to J/\psi \KL$ analysis~\cite{macfprd}
either as a cluster of energy deposited in the EMC
or as a cluster of hits in two or more layers of the IFR
that cannot be associated with any charged track in the event.
The \KL energy is not measured, therefore, we determine the \KL
laboratory momentum from its flight direction as measured from the EMC or IFR
cluster, and the constraint that the invariant \kkkl\ mass agree with the
known $B^0$ mass. In those cases where the \KL is detected in both the IFR and EMC
we use the angular information from the EMC, because it has higher precision.
In order to reduce background from $\pi^0$ decays, we reject
an EMC \KL\ candidate cluster if it forms an invariant mass between 100
and 150~\mevcc with any other neutral cluster in the event under the $\gamma\gamma$
hypothesis, or if it has energy greater than 1~GeV and contains two shower
maxima consistent with two photons from a $\pi^0$ decay.
The remaining background of \KL candidates due
to photons and overlapping showers is further reduced
with the use of a neural network ${\cal NN}_{EMC}$.
The ${\cal NN}_{EMC}$ is constructed from cluster shape variables,
trained using as signal measured $B^0\to J/\psi \KL$ events and 
as background measured \kkkl\ events which lie outside the signal region, 
and tested on $e^+e^-\to\phi(\to\KS\KL )\gamma$ events.

The results are extracted using an extended unbinned maximum likelihood fit.
We parameterize the distributions of
kinematic and topological variables for signal and background events
in terms of probability density functions (PDFs)~\cite{oldpub}.
The selection requirements for these variables is loose to
include background dominated regions which can then be 
extrapolated  into the signal region.
The main source of background, estimated from data,
comes from random combinations of tracks
produced in events of the type $e^+e^-\to q\bar{q}$, where
$q = u,d,s,c$ (continuum).
Background from decays of $B$ mesons in other final states
with and without charm is estimated using Monte Carlo simulation.

In the following we describe the event variables
used in the maximum likelihood fit to characterize the signal and background:
the energy difference $\Delta E = E_B^* - \frac{1}{2}\sqrt{s}$, where
$E_B^*$ is the energy of the \B\ candidate and $\sqrt{s}$ is the 
total energy, both evaluated in the $\Upsilon(4S)$ rest frame, 
a neural network $\cal NN$ built with topological quantities,
and \deltat, described in Section~\ref{sec:Introduction}.
For signal events, \DeltaE\ is expected to peak at zero, with a 
broad tail for positive values of \DeltaE.  We require
$\Delta E < 0.08$~GeV, in order to be able to fix the shape of background under the signal peak.
The \DeltaE\ resolution is 3.0 \mev, which includes the different resolutions 
for EMC and IFR events. This resolution has been validated on data using reconstructed \Bztojpsikl\ events.

Continuum events are
characterized by a jet-like topology in the $\Upsilon(4S)$ rest frame, 
because of the large amount of phase space in the decay,
while $B$ mesons are produced almost at rest, so particles produced in 
$B$ decays are distributed isotropically.
One can then define a set of topological variables to quantify the sphericity of the event.
One such quantity is the angle $\theta_{SPH}$ between the sphericity axis of the
$B_{CP}$ candidate and the sphericity axis formed from the other charged and
neutral particles in the event. 
We also use the cosine of the angle $\theta_B$ between the $B_{CP}$ momentum and
the beam axis, and the sum of the momenta $p_i$ of the other charged and neutral
particles in the event weighted by the Legendre polynomials $L_0(\theta_i)$
and $L_2(\theta_i)$ where $\theta_i$ is the angle between the momentum of
particle $i$ and the thrust axis of the $B_{CP}$ candidate.
We use other variables characterizing a final state with a \KL. 
One is the reconstructed  energy difference $\Delta E_{vis}$, 
calculated in the laboratory frame as the difference between total
reconstructed energy of the event,
where the energy of the neutral particles is 
calibrated for an electromagnetic shower,
and the two reconstructed kaon energies.
The other is the cosine of the polar angle of the 
missing momentum, $\vec{p}_{miss}$, calculated as 
the difference between the sum of beam momenta
and all tracks and EMC clusters, in the laboratory frame, excluding the \KL candidate.
We combine these variables in a neural network $\cal NN$, which 
peaks at 0 for continuum events and at 1 for signal events. We apply a selection 
that retains $80\%$ of signal events and  rejects 
$84\%$ of continuum background. The rest of the events are used for the maximum likelihood fit.

The remaining background originates from \B\ decays where a neutral or charged pion is 
missed during the reconstruction (peaking \B\ background). Since the branching fractions for 
these decay modes ($B^0\to K^+K^-K^{*0}(K^0_L\pi^0)$, $B^0\to K^{*+}(K^+\pi^0)K^-K^0_L$, 
$B^+\to K^+K^-K^{*+}(K^0_L\pi^+)$, and $B^+\to K^+K^0_LK^{*-}(K^-\pi^0)$) are not known, we build a cocktail
of exclusive Monte Carlo samples, weighted with the relative efficiency, and the yield is floated
in the final fit. 
The rest of the background originating from \B\ decays comes from the combinations of particles originating 
from both $B$ mesons that have continuum-like values of \DeltaE, 
so  these events are included in the continuum 
component by the fit, without generating a bias in the fitted values of $S$ and $C$ 
parameters\footnote{This conclusion has been obtained simulating the fit on a sample of 
fully simulated Monte Carlo events.}.

We suppress background from $B$ decays that proceed through a $b \to
c$ transition leading to the \kkkl\ final state by applying invariant
mass cuts to remove $\Dz$, $\jpsi$, $\chi_{c0}$, and $\psi(2S)$
decaying into $\Kp\Km$,\ and $\Dp$ and $\Ds$ decays into
$\Kp\KL$.

All the other tracks and clusters that are not associated with the
reconstructed \Bztokkkl\ decay are used to form the \Btag;
its flavor is determined with a multivariate tagging algorithm~\cite{jpsinew}.
The tagging efficiency $\varepsilon$ and mistag probability $w$ in five hierarchical and
mutually exclusive categories are measured using fully reconstructed
$B^0$ decays into the $D^{(*)-}X^+\,(X^+ = \pip, \rho^+, a_1^+)$  and
$\jpsi K^{*0}\,(K^{*0}\to\Kp\pim)$ flavor eigenstates (\Bflav sample).
The analyzing power $\varepsilon (1-2w)^2$ is $(30.3\pm 0.4)$\%.

A detailed description of the $\Delta t$ reconstruction algorithm is
given in Ref.~\cite{macfprd}.

\section{MAXIMUM LIKELIHOOD FIT}
\label{sec:MlFit}
The \CP\ asymmetry parameters are extracted from a \kkkl\ sample which excludes \phikl\ 
with an invariant mass veto: $|m(K^+K^-)-m(\phi)|> 15 \mevcc$.
This excludes \phikl\ events by three standard deviations.
The average $\deltaz$ resolution is $190\mu m$,
dominated by the tagging vertex in the event.
Thus, we can characterize the resolution using the much larger
$B_{\mbox{flav}}$ sample, which we use for signal parameterization.
The amplitudes for the $B_{CP}$ asymmetries and for the $B_{\mbox{flav}}$
flavor oscillations are reduced by the same factor due to
wrong tags. Both distributions are convolved with a common $\Delta t$
resolution function. Backgrounds are accounted for by adding
terms to the likelihood, incorporated with different assumptions
about their $\Delta t$ evolution and resolution function~\cite{macfprd}.
The \deltat\ resolution function is parameterized as a sum of two Gaussian distributions
with different mean values
whose widths are given by a scale factor times the event-by-event uncertainty $\sigma_{\deltat}$.
A third Gaussian distribution, with a fixed large width, accounts for a small
fraction of outlying events~\cite{sin2bnewbabar}.
For the time-dependent fit we retain events that have $|\Delta  t|<$ 20 ps and whose estimated uncertainty 
$\sigma_{\Delta t}$ is less than 2.5 ps. 

\begin{figure}[ht]
\begin{center}
\begin{tabular}{cc}
\epsfig{file=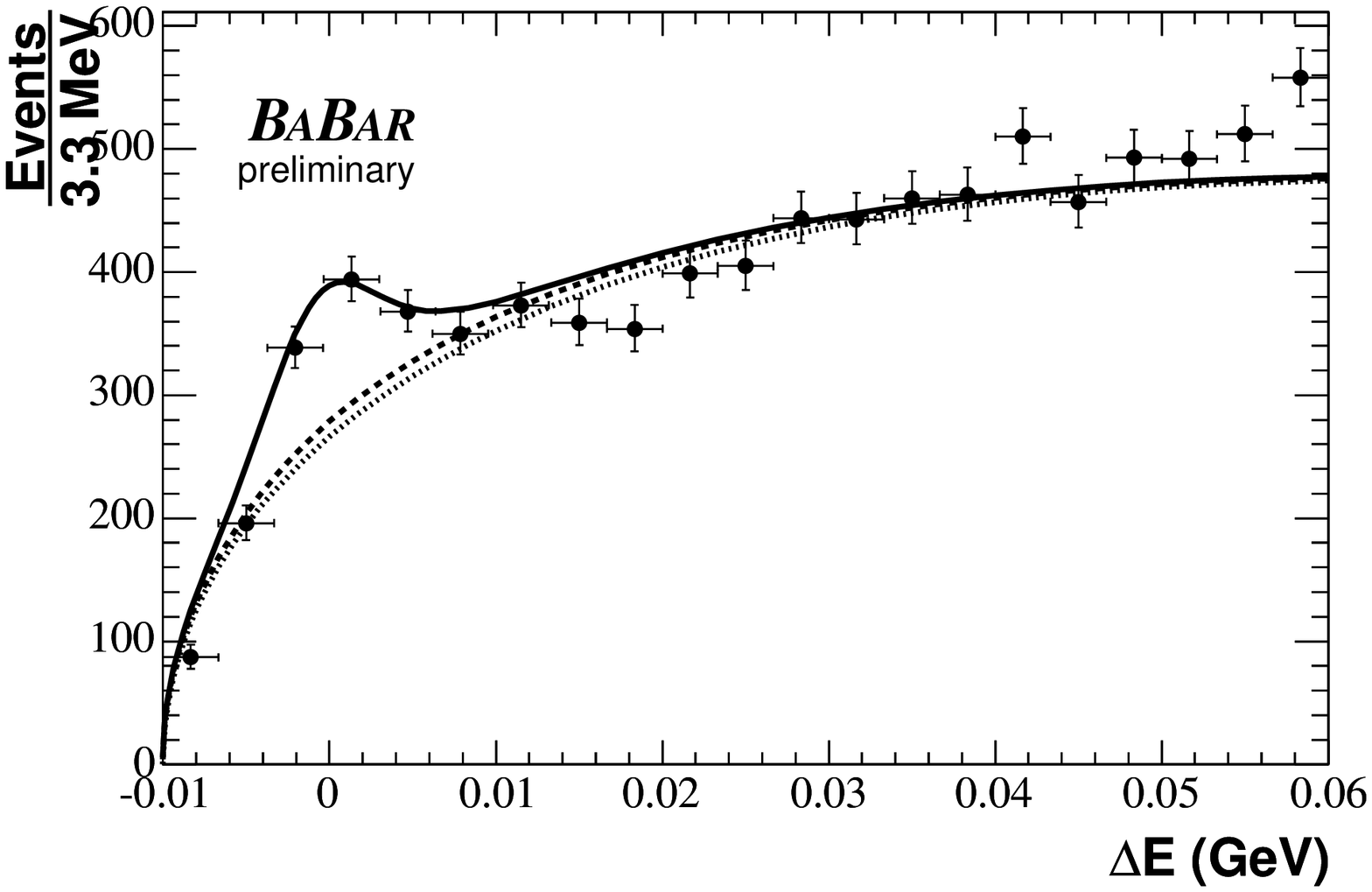,width=8.cm}
\epsfig{file=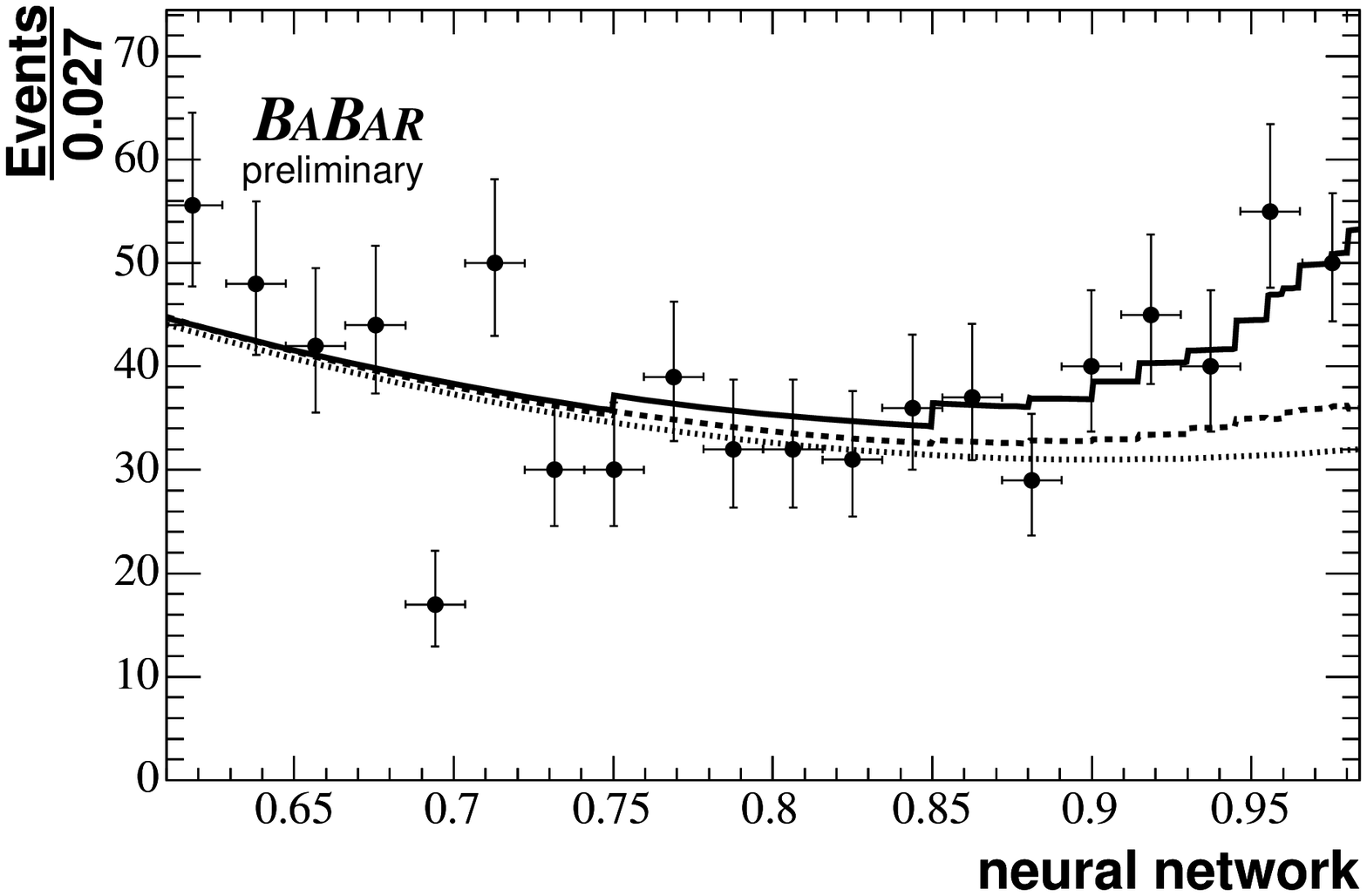,width=8.cm}
\end{tabular}
\caption{\label{fig:denn} Distribution of the event variable \DeltaE\ 
after a $\cos\theta_H$ cut (left)
and neural network output after  a requirement on the likelihood,
calculated without the plotted variable.
The signal efficiency for the selection and likelihood requirements
is 31\% for \DeltaE\ and 8\% for the neural network output.
The solid line represents the fit result
for the total event yield and the dashed line for the total
background. The dotted line represents the continuum
background, only.}
\end{center}
\end{figure}

\begin{figure}[ht]
\begin{center}
\begin{tabular}{cc}
\epsfig{file=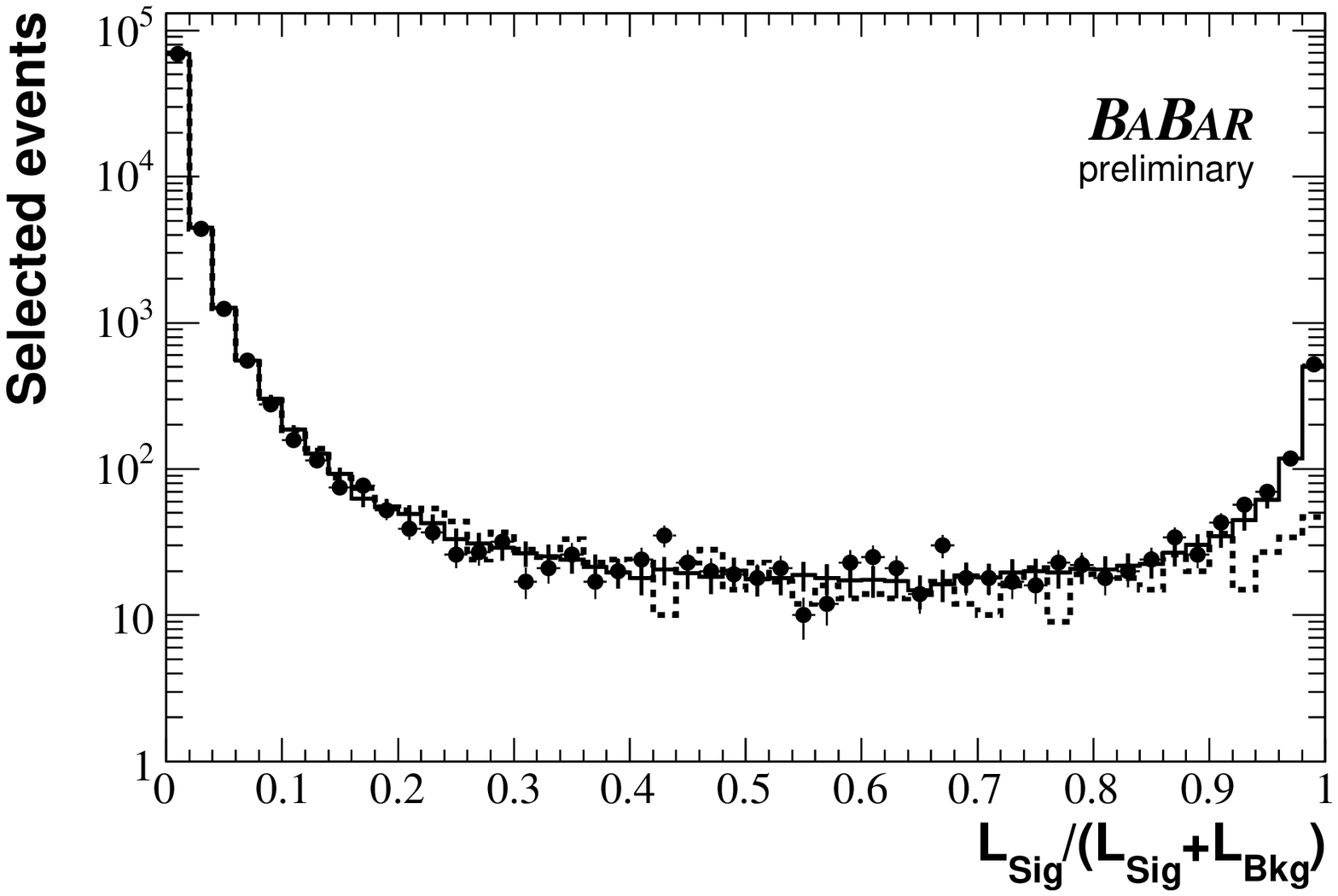,width=8.cm}

\end{tabular}
\caption{\label{fig:LsNorm} 
Distribution of signal to background likelihood ratio
$L_{\mbox{sig}}/(L_{\mbox{sig}}+L_{\mbox{bkg}})$. The solid line represents the 
Monte Carlo simulation of the entire sample, and the dashed line the 
simulation of background only.
}
\end{center}
\end{figure}

Since we measure the correlation among the observables to be small in the data sample used in 
the fit (the largest is 2.9 \% between $\Delta E$ and $\cal NN$) we take the probability density
function $\mathcal{P}_{i,c}^j$ for each event $j$ to be a product of the
PDFs for the separate observables. For each event hypothesis $i$
(signal, backgrounds) and tagging category $c$, we define
$\mathcal{P}_{i,c}^j =\mathcal{P}_i(\Delta E)
\cdot\mathcal{P}_i(\mathcal{NN})
\cdot\mathcal{P}_i(\Delta t;\sigma_{\Delta t}, c)$,
The likelihood function for each decay chain is then
\begin{equation}
\label{likel}
{\mathcal L} = \prod_{c}\exp{\left(-\sum_{i}N_{i,c}\right)}
\prod_{j}^{N_c}\left[\sum_{i}N_{i,c}\,{\mathcal P}_{i,c}^j \right] ,
\end{equation}
where $N_{i,c}$ is the yield of events of hypothesis $i$ obtained from the
fit in category $c$, and $N_c$ is the number of category $c$ events in the sample.
The total sample consists of 77577 \kkkl\ candidates. The total reconstruction efficiency is
$\langle \varepsilon  \rangle =(23.1 \pm 0.6)\%$.
We fixed in the fit  $S_{B-bkg}=0.42$ (as estimated from full Monte Carlo simulation
of generic neutral and charged \B-decays)
and $C_{B-bkg}=0$ (as the SM expectations). 
From the fit we find 777 $\pm$ 80 \kkkl\ signal events (\phikl\ excluded). 
The signal yield agrees with the branching fraction determined in the \kkks\ final state
within one standard deviation,
but the uncertainty in \KL\ efficiency is large.
Figure \ref{fig:denn} shows the \DeltaE\ distribution together with 
the result from the fit after a $\cos \theta_H$ 
cut to enhance signal, where $\theta_H$ is the angle between the $K^+$ candidate and the parent
$B_{CP}$ flight direction in the $K^+K^-$ rest frame.
The neural network output distribution together with 
the result from the fit after a requirement on the likelihood to enhance the sensitivity.
The fit was tested with both a parameterized simulation of a large
number of data-sized experiments and a full detector simulation.
The likelihood of our data fit agrees with the likelihoods from
fits to the simulated data.
Figure \ref{fig:LsNorm} shows the comparison between data and Monte Carlo
simulated events of the signal to background likelihood ratio
$L_{\mbox{sig}}/(L_{\mbox{sig}}+L_{\mbox{bkg}})$ distribution.
It shows the goodness of the agreement between data and
Monte Carlo parameterization event-by-event.
The fit was also verified with our $J/\psi \KL$ data sample to check 
the fitted central value of \stwob\ and the sign of the \CP\ eigenstate definition. 

\section{ESTIMATION OF \CP\ CONTENT}
\label{sec:cpcont}
The measurement of the \CP\ content has been done in the \kkks\ final state from
an angular moments analysis~\cite{kkksconf}. 
Since the \kkks\ final state has higher purity than can be obtained in the \KL\ 
final state, the angular analysis has not been repeated with our sample,
but the results on the \CP-conjugate state have been used. 
In order to take into account differences in the efficiency across the Dalitz plot 
between the \kkks\ and \kkkl\ samples, which
can change the relative amount of \CP-even(odd) fraction in different $m(K^+K^-)$ 
regions,
we use the $f_{even}$ fraction measured in seven $m(K^+K^-)$ bins 
(excluding the $\phi$ region) in 
\Bztokkks\ and compute a re-weighted average using \kkkl\ yields in the same mass bins.
Table \ref{tab:feven} shows the \kkkl\ efficiencies, yields and the measured \kkks\ $f_{even}$.
Variations of the signal yield are shown also in the upper plot of Fig. \ref{fig:sub}.

\begin{table}[!t]
\begin{center}
\caption[.]{\label{tab:feven} Average efficiencies, yields for $B^0\to K^+K^-K^0_L$, and $f_{even}$, 
calculated in the $K^+K^-K^0_S$ final state, for seven $K^+K^-$ mass bins (excluding $\phi$ region).}
\vspace{0.1cm}
\setlength{\tabcolsep}{1.12pc}
\begin{tabular}{cccc}
\hline\hline
&&& \\[-0.3cm]
$m(K^+K^-)$ \gevcc   & $\langle \varepsilon  \rangle $   & $Signal~yield$                & $f_{even}$ (\kkks\	) \\
&&& \\[-0.3cm]
\hline\hline
$[1.1~;~1.3]$           & 0.153                         & 67.6 $\pm$ 20.8      & 1.10 $\pm$ 0.18         \\
$[1.3~;~1.5]$           & 0.172                         & 44.0 $\pm$ 21.3      & 0.99 $\pm$ 0.14        \\
$[1.5~;~1.9]$           & 0.188                         & 93.9 $\pm$ 28.0      & 0.93 $\pm$ 0.21       \\
$[1.9~;~2.3]$           & 0.223                         & 146.4 $\pm$ 28.1     & 0.95 $\pm$ 0.16        \\
$[2.3~;~2.7]$           & 0.258                         & 117.4 $\pm$ 24.0     & 0.79 $\pm$ 0.24        \\
$[2.7~;~3.1]$           & 0.271                         & 93.7 $\pm$ 21.8      & 0.74 $\pm$ 0.25        \\
$[3.1~;~4.9]$           & 0.242                         & 141.9 $\pm$ 37.2     & 0.96 $\pm$ 0.43        \\
\hline\hline
\end{tabular}
\end{center}
\end{table}
Out of the $\phi$ region the sample  consists mainly of
S-wave decays, giving an $f_{odd}$ fraction close to 1. We find the 
total fraction of \CP-odd final states:
$$
        f_{odd} ~ = ~  0.92 \pm 0.07~(\mbox{stat}) \pm 0.06~(\mbox{syst}) ,
$$
where the systematic uncertainty is evaluated in the \babar\ \Bztokkks\ analysis~\cite{kkksconf}.

\section{SYSTEMATIC STUDIES}
\label{sec:Systematics}
We consider systematic uncertainties in the \CP\ coefficients $S$ and $C$ due to 
the parameterization of PDFs for the event yield in signal and background 
by varying the parameters within one standard deviation
(evaluated from a fit to Monte Carlo simulated events).
Since the real \CP content of the \B\ background is not known, we 
vary $S_{B-bkg}$ and $C_{B-bkg}$ in a conservative interval.
About 50\% of these events come from charged \B\ decays, which show 
only direct \CP-violation, while neutral \B\ decays can violate \CP\ both
in mixing and decay. 
We therefore vary the \B\ background \CP\ parameters in the interval 
$-0.5 < S_{B-bkg} (C_{B-bkg}) < 0.5$, which corresponds to a uniform variation of the 
parameters for neutral \B\ background in the whole 
physically allowed interval
$-1 < S_{B-bkg} (C_{B-bkg}) < 1$.
We evaluate the uncertainty associated with
the assumed parameterization of the \deltat\  resolution function for signal 
and \B\ background, a possible difference in the efficiency between 
reconstructed \Bz\ and \Bzb\ decays, and the fixed values 
for \deltamd\ and $\tau_{B^0}$, by varying the parameters within one standard deviation
(extracted from a fit to the \Bflav\ sample).
We also estimate uncertainties coming from possible 
SVT layer misalignments.
The bias in the coefficients due to the fit procedure
is included in the uncertainty without making corrections to the final results.
Finally, we estimate the errors due to the effect of doubly
CKM-suppressed decays~\cite{Long:2003wq}.
We add these contributions in quadrature to obtain the total
systematic uncertainty. The summary is reported in Table \ref{tab:syst}.
\begin{table}[hbtp]
\begin{center}
\caption{\label{tab:syst}Summary of systematic uncertainties on the
parameters $S$ and $C$.
The total systematic errors are obtained by adding in quadrature all individual sources.}
\begin{tabular}{lcccc} \\ \hline\hline
&&&&\\[-0.3cm]
Source & $\Delta~S(+)$ & $\Delta~S(-)$ & $\Delta~C(+)$ & $\Delta~C (-)$  \\
&&&&\\[-0.3cm]
\hline
$\Delta m_d$                            & 0.004 & -0.001 & 0.000 & -0.001 \\
$\tau_{B^0}$                    & 0.01 & 0.01 & 0.00 & -0.00 \\
$\Delta t$ model         & 0.02       & 0.02 & 0.02 & 0.01    \\
Tagging & 0.04       & 0.04 & 0.03 & 0.03   \\
\B background \CP\                & 0.10 & 0.10 & 0.06 & 0.07 \\
Signal and background PDFs               & 0.03       & 0.03 & 0.02 & 0.02   \\
fit biases                & 0.00         & 0.00  & 0.02 & 0.02 \\
SVT local alignment             & 0.01        & 0.01        & 0.00        & 0.00        \\
doubly-CKM-suppressed decays    & 0.00        & 0.00        & 0.01         & 0.01 \\
\hline
Total                   & 0.11 & 0.12 & 0.08 & 0.09 \\
\hline\hline
\end{tabular}
\end{center}
\end{table}

\section{RESULTS}
\label{sec:Physics}
\begin{figure}[hpt]
\begin{center}
\vskip 1cm
\epsfig{file=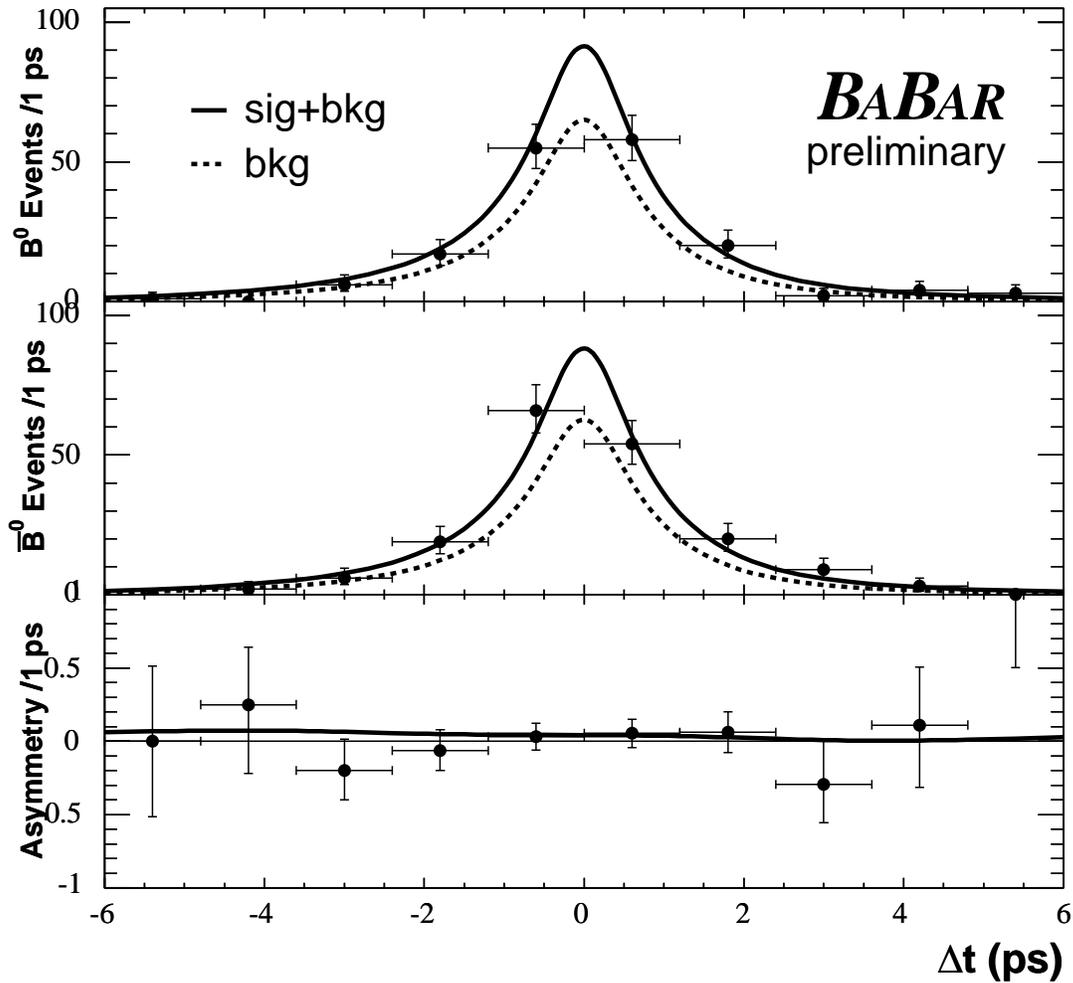, width=15cm}
\caption{Distributions of \deltat\ for \Bztokkkl\ candidates with (top) \Bz- and (middle) \Bzb-tags.
The solid lines refer to the fit for all events; the dashed lines
correspond to the background. The bottom plot shows the asymmetry.
A requirement on signal-to-background ratio to enhance the signal is applied.
\label{fig:b0b0bar}}
\vspace*{-0.5cm}
\end{center}
\end{figure}

\begin{figure}[hpt]
\begin{center}
\vskip 1cm
\epsfig{file=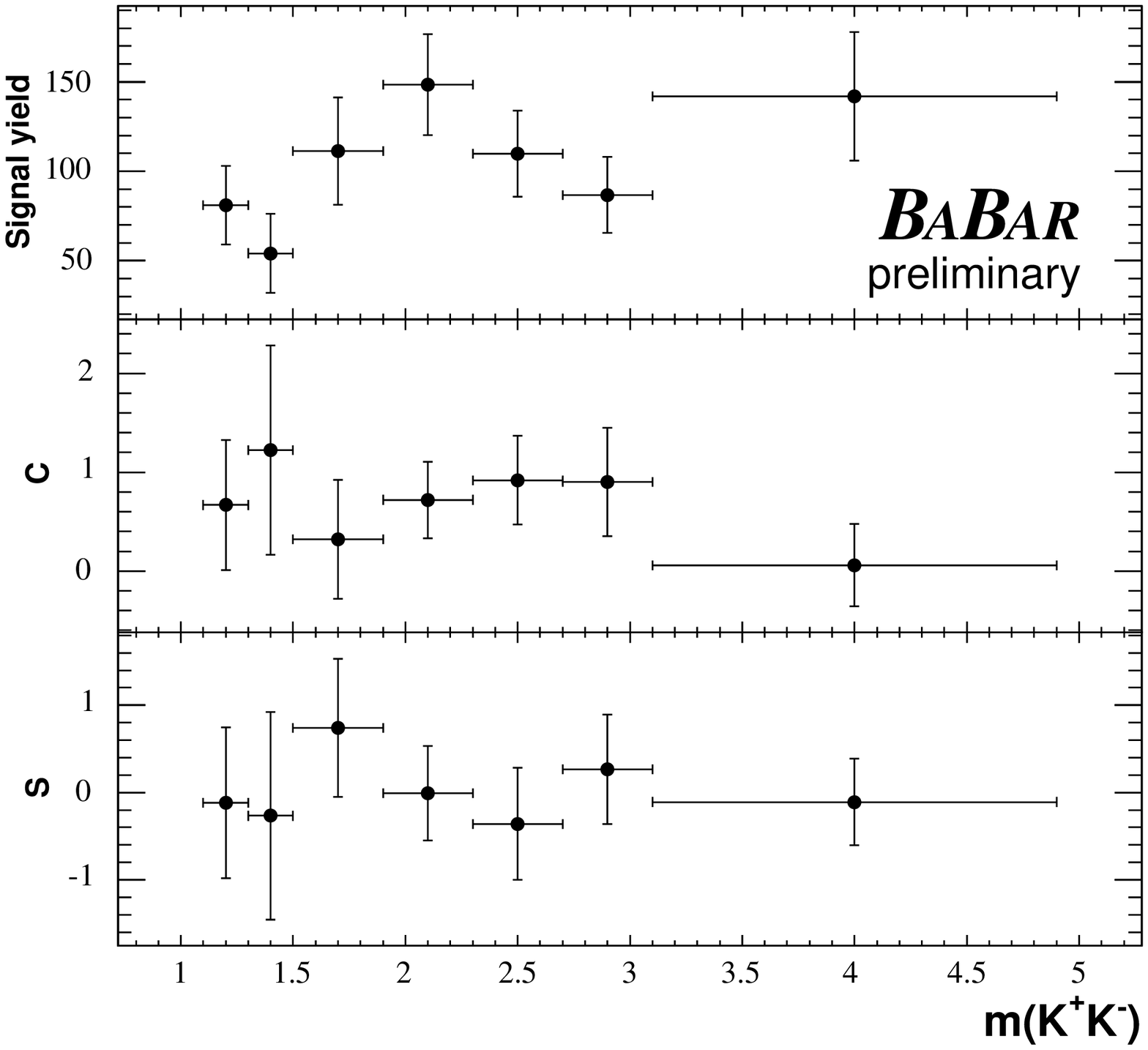, width=15cm}
\caption{Distribution of signal yield (top), $C$ (middle) and $S$ (bottom) 
parameters in 7 different $m(K^+K^-)$ intervals.
\label{fig:sub}}
\vspace*{-0.5cm}
\end{center}
\end{figure}

The coefficients of the time-dependent \CP\ asymmetry in \Bztokkkl\ decays (excluding the \phikl\ 
final state) determined by the fit are:
\begin{eqnarray}
S & = & \finalscb , \nonumber \\
C & = & \finalccb . \nonumber
\end{eqnarray}
Figure \ref{fig:b0b0bar} 
shows the $\deltat$ distributions of the \Bz- and the \Bzb-tagged subsets together with the raw asymmetry, 
with the result of the combined time-dependent \CP-asymmetry fit superimposed.
We also fit the \CP\ parameters in the same $m(K^+K^-)$ regions used to estimate the \CP-odd fraction.
The results are shown in Figure \ref{fig:sub}.
The presence of both P- and  S-wave decays in our \CP\ sample dilutes the measurement of the sine coefficient.
If we account for the measured \CP -odd fraction, we can extract the SM parameter $\sin 2\beta$.
Using the estimate of the \CP\ content based on the angular moments, and setting $C=0$ in the fit, we get
$$
\stwob~_{\mbox{eff}} ~=~ S/(2f_{odd}-1)~=~\finalstwob
$$
where the last error is due to uncertainty on the \CP\ content. 
Since this uncertainty is multiplicative and the fitted value of $S$ is 
close to 0, we conservatively computed this uncertainty shifting the measured value of $S$ within 1 
standard deviation.

\section{SUMMARY}
\label{sec:Summary}
In a sample of 227 million \BB\ mesons, we have obtained preliminary measurements of the \CP\ content
and \CP\ parameters  in the \kkkl\ final state that excludes \phikl\ decays.
We estimated the fraction of S-wave events (\CP-odd fraction) 
from measured value in the \kkks\ final state,
which has higher purity than \kkkl. The result shows the dominance of \CP-odd final states.
We compute the average of $\stwob_{\mbox{eff}}$ and $C$ parameter in \kkks\ and \kkkl\ final states 
treating the systematic errors and the uncertainty on the \CP content as completely correlated.
This gives the most conservative estimation of the combined uncertainty. 
We obtain: 
\begin{eqnarray*}
\left[ \stwob~_{\mbox{eff}} \right]_{\mbox{av}} ~&=&~0.41 \pm 0.18~(\mbox{stat}) \pm 0.07~(\mbox{syst}) \pm 0.11~(\CP \mbox{content}) \\
\left[ C \right]_{\mbox{av}} ~&=&~0.23 \pm 0.12~(\mbox{stat}) \pm 0.07~(\mbox{syst})\\
\end{eqnarray*}
The result agrees within one standard deviation with the value of \stwob in the
$B^0\to (\bar{c}c) K^0$ decays~\cite{jpsinew}.

\section{ACKNOWLEDGMENTS}
\label{sec:Acknowledgments}

We are grateful for the 
extraordinary contributions of our \pep2\ colleagues in
achieving the excellent luminosity and machine conditions
that have made this work possible.
The success of this project also relies critically on the 
expertise and dedication of the computing organizations that 
support \babar.
The collaborating institutions wish to thank 
SLAC for its support and the kind hospitality extended to them. 
This work is supported by the
US Department of Energy
and National Science Foundation, the
Natural Sciences and Engineering Research Council (Canada),
Institute of High Energy Physics (China), the
Commissariat \`a l'Energie Atomique and
Institut National de Physique Nucl\'eaire et de Physique des Particules
(France), the
Bundesministerium f\"ur Bildung und Forschung and
Deutsche Forschungsgemeinschaft
(Germany), the
Istituto Nazionale di Fisica Nucleare (Italy),
the Foundation for Fundamental Research on Matter (The Netherlands),
the Research Council of Norway, the
Ministry of Science and Technology of the Russian Federation, and the
Particle Physics and Astronomy Research Council (United Kingdom). 
Individuals have received support from 
CONACyT (Mexico),
the A. P. Sloan Foundation, 
the Research Corporation,
and the Alexander von Humboldt Foundation.


\begin{thebibliography}{99}

\bibitem{ckm}
N.~Cabibbo, \jprl {\bfseries 10}, 531 (1963);
M.~Ko\-ba\-ya\-shi and T.~Maskawa, \progtp {\bfseries 49}, 652 (1973).

\bibitem{one}
N.~G.~Deshpande and J.~Trampetic, \jprd{41}, 895 (1990);
N.~G.~Deshpande and G.~He, \plb{336}, 471 (1994);
R.~Fleischer, Z.\ Phys.\ C {\bfseries 62}, 81 (1994);
Y.~Grossman {\itshape et al.}, \jprd{68}, 015004 (2003).

\bibitem{grossman}
A.~B.~Carter and A.~I.~Sanda, \jprd{23}, 1567 (1981);
I.~I.~Bigi and A.~I.~Sanda, \npb{193}, 85 (1981);
Y.~Grossman and M.~P.~Worah, \plb{395}, 241 (1997);
R.~Fleischer, Int.\ J.\ Mod.\ Phys.\ A {\bfseries 12}, 2459 (1997);
D.~London and A.~Soni, \plb{407}, 61 (1997).

\bibitem{Ciuchini:2002uv}
  M.~Ciuchini, E.~Franco, A.~Masiero and L.~Silvestrini,
  Phys.\ Rev.\ D {\bf 67}, 075016 (2003)
  [Erratum-ibid.\ D {\bf 68}, 079901 (2003)]
  [arXiv:hep-ph/0212397].

\bibitem{sin2bnewbabar}
B.~Aubert {\itshape et al.}, [\babar\ Collaboration], \jprl{89}, 201802 (2002).

\bibitem{sin2bnewbelle}
K.~Abe {\itshape et al.}, [Belle Collaboration], \jprd{66}, 071102 (2002).

\bibitem{pdg}
S. Eidelman {\itshape et al.}, [Particle Data Group], Phys. Lett. {\bf B} 592, 1 (2004).

\bibitem{fits}
J.~Charles {\it et al.}  [CKMfitter Group], arXiv:hep-ph/0406184;
M.~Bona {\it et al.}  [UTfit Collaboration], arXiv:hep-ph/0501199.

\bibitem{Abe:2003yt}
K.~Abe {\it et al.}  [Belle Collaboration],
Phys.\ Rev.\ Lett.\  {\bf 91}, 261602 (2003).

\bibitem{Aubert:2004ii}
B.~Aubert  [\babar\ Collaboration],
BABAR-CONF-04/033

\bibitem{kkksconf}
B.~Aubert  [\babar\ Collaboration],
BABAR-CONF-04/025

\bibitem{Aubert:2001tu}
B.~Aubert {\itshape et al.}, [\babar\ Collaboration], \nima{479}, 1 (2002).

\bibitem{pep}
PEP-II Conceptual Design Report, SLAC-R-418 (1993).

\bibitem{macfprd}
B.~Aubert {\itshape et al.}, [\babar\ Collaboration], \jprd{66}, 032003 (2002).

\bibitem{jpsinew}
B.~Aubert {\itshape et al.}, [\babar\ Collaboration],
`Improved Measurement of Time-Dependent \CP Violation in
$B^0\to (\bar{c}c) K^0$ Decays', Contribution to ICHEP~2004.

\bibitem{Long:2003wq}
O.~Long, M.~Baak, R.~N.~Cahn and D.~Kirkby, \jprd{68}, 034010 (2003).

\bibitem{oldpub}
B.~Aubert {\itshape et al.}, [\babar\ Collaboration], \jprl{87}, 151801 (2001).






\end{thebibliography}
\end{document}